\documentstyle[aas2pp4]{article}

\newcommand\etal{et al.}
\newcommand\secpoint{\mbox{$''\mskip-7.6mu.\,$}}

\journalid{337}{15 January 1989}
\articleid{11}{14}

\begin{document}

\lefthead{Visvanathan \& Wills}
\righthead{Blazar Polarization}
\title{Optical Polarization of 52 Radio-Loud QSOs and BL Lac Objects}

\author{Natarajan Visvanathan}
\affil{Mount Stromlo \& Siding Spring Observatories, Australian National University, ACT 2611,
Australia; vis@mso.anu.edu.au}
\and
\author{Beverley J. Wills}
\affil{Astronomy Department \& McDonald Observatory, University of Texas, Austin, TX 78712, USA;
bev@astro.as.utexas.edu}

\begin{abstract}

Polarization measurements are presented for 52 radio-loud QSOs and BL Lac objects.
For 9 highly polarized (p$>3$\%) AGN, these are the first published polarization 
measurements.  Of these 9, 7 are highly-polarized QSOs (HPQs), one is a BL Lac object and
another is a likely BL Lac object.
Polarization variability is confirmed for some of these new and previously known
highly-polarized AGN.  While 6 of the HPQs have flat radio spectra are almost certainly
blazars, PKS\,1452$-$217 is probably a new member of the rare class of 
radio-loud QSOs that show high polarization by scattering, and is therefore 
important for testing orientation Unified Schemes.
In competition for the highest redshift HPQ are the well-observed QSO PKS\,0438$-$43 at 
z$ = $2.85, with maximum p$\sim$4.7\%, and PKS\,0046$-$315 at z$= 2.72$, for which we find
p$\sim$13\%.

\end{abstract}




\vfil
\vfil\eject
\section{Introduction}

The optical polarization of quasars and the even more violently variable BL Lac objects 
is of interest for understanding the origin, confinement and propagation of relativistic
jets (e.g. \cite{Br85}; \cite{Si97}; \cite{Wa96}; Gabuzda, Sitko, \& Smith 1996).
Thus there have been many attempts to relate the polarization and its variability
to other properties of the optical spectra (e.g. \cite{SMA84};
\cite{Me90}), to radio spectra (e.g. \cite{MS84}), and to radio structure, 
especially in relation to orientation and Unified Schemes (e.g. \cite{Ru90};
\cite{AU85}). 

The polarization of highly polarized QSOs and BL Lac objects (collectively known as
blazars) has been extensively
monitored on time scales of days to months (e.g. \cite{Mo82}; \cite{Ki95}).
Here we present polarization measurements made in the period 1978 -- 1979, which will
contribute to studies of longer time-scale variability (\cite{Tr98}), as well
as providing new data on little-studied sources.
Most of the sample consists of originally suggested identifications for
Parkes Survey (PKS) radio sources.  They were or have been
since confirmed as quasars.  Two non-PKS sources, OX\,169 and OJ\,287 were included,
sinply to see if they were polarized, or to search for variable polarization.

\section{Observations and Results}

All the observations were made at the Cassegrain focus of the Anglo
Australian Telescope using the automated polarimeter (\cite{Vi72}).
The polarimeter uses a rotating HNP$^{\prime}$B (Polaroid, UV) linear polarizer
as the analyser.  To avoid apparent polarization caused by polarization 
sensitivity of succeeding optical elements, the polarizer is followed by a fixed
calcite depolarizer mounted in front of the aperture wheel. The
aperture wheel is followed by a star-sky chopper that operates with a
frequency of 30\,Hz and which is used in conjunction with two equal-sized apertures
in the focal plane, in order to reduce the effects of short-term fluctuations in the
polarized night sky.  No band pass filter was used, and the detector was an S20
photomultiplier.

The polarizer is rotated by a stepping motor, each step corresponding to 2$^\circ$ in
polarization position angle.
The chopper is synchronized  with the polarizer movement such that for
every position of the polarizer one star and one sky  exposure is made.  A
full rotation of the polarizer through 360$^\circ$\ takes 12 seconds. To obtain better
statistics the observations are continued for 4, 12, or 24 rotations.
The diameters of the  apertures used were 3\secpoint6 and 8\secpoint1.  The
counts from
the star and sky are displayed, and polarization is detected by the appearance
of a double sinusoid.
At the end of the observations the sky signal is subtracted from the combined
object-plus-sky signal, and
polarization, p, and position angle, $\theta$, are evaluated using a FFT program.

The instrumental polarization of the telescope was derived by observing
unpolarized stars and was found to be small -- 0.46\% $\pm$ 0.04\% at 136$^\circ$.  
The position angle measured from
north through east has been calibrated by observing stars of known polarization.

The results are given in Table\,1. 
Column 1 gives the B1950 co-ordinate
name.  Column
2 gives the radio survey name from the NASA Extragalactic Database (NED\footnote{ 
http://adsabs.harvard.edu/abstract\_service.html}).  Where only `PKS' is given, the 
survey name can be formed by appending the co-ordinate name of column 1.
If the optical object came from a non-PKS reference, that name is given.  Other
well-known names are given in column 3.
Column 4 gives the object classification as QSO or BL Lac object (BLL), generally
based on the information in the V\'eron-Cetty \& V\'eron (1993) catalog, or otherwise
in NED, except for PKS\,B1921$-$293 (OV$-$236), which should be classified as a 
BL Lac object (\cite{Wi81}).
`HP' for `high polarization' means that polarization p\,$>$\,3\% has been reported
in the literature (\cite{SF97}; \cite{St94}; \cite{IT90};
\cite{Wi92}).  Column 5 gives the redshift, where available.  Column 6 gives
the radio spectral index, $\alpha$\ (F$_{\nu} \propto {\nu}^{-\alpha}$),
determined between 2.7\,GHz and 5\,GHz.  Columns 7 -- 9 give the UT date of 
observation, the percentage polarization, p, with rms uncertainty
$\sigma_{\rm p}$, and the polarization position angle, $\theta$.  Note that, for
$\sigma_{\rm p} <<$\ p, the rms uncertainty in $\theta$\ can be estimated from
$ 28.65 \times {\sigma_{\rm p}}$/p.  
Correction for instrumental polarization has been made, but
no correction has been applied for positive bias
in p that arises from the derivation p = $({\rm Q}^2 + {\rm U}^2)^{1/2}$.  For
cases of large rms uncertainties $\sigma_{\rm p}$ compared with p, a
better estimate of the true polarization is 
$({\rm p}^2 - {\sigma_{\rm p}}^2)^{1/2}$ (\cite{WK74}).  For comparison
with our p in column 8, in column 10 we give the expected maximum contribution from 
interstellar polarization in our Galaxy, isp, derived from 9\% $\times E(B-V)$\ 
(\cite{SMF75}).  A likely value of interstellar polarization
can be derived from $\sim$4.5\% $\times E(B-V)$, although the uncertainties are large
(and the polarization position angle is not specified).  We use values of
$E(B-V)$ derived from $A_{\rm B}/4.1$\ where $A_{\rm B}$ is given in NED, thanks
to Burstein \& Heiles (1982).  In assessing the reality of
polarization intrinsic to the AGN, the positive bias should be considered,
together with the possible contribution of Galactic interstellar polarization.

\section{ Discussion}

Table\,1 
gives the first published polarization measurements for 9 highly 
polarized
(p\,$>$\,3\%) objects: 7 QSOs (PKS\,0046$-$315, 0422$-$380, 1034$-$374, 1207$-$399,
1216$-$010, 
1256$-$22, \& 1452$-$217), one BL Lac object (PKS\,1307+12C), and another, apparently
classified as a galaxy but likely to be a BL Lac object (PKS\,B1206$-$238).
In addition, the other measurements, in combination with literature data, extend the
time coverage, as well as providing two or more measurements that give evidence for
time-variability and hence confirmation of the synchrotron nature of the polarization.
From our data, variable polarization is seen in PKS\,0754+100 (OI+090.4), 0823$-$223,
0829+046, 1921$-$293 (OV$-$236), and PKS\,2155$-$304.
Some objects have low but significant polarization that is likely to be intrinsic to the
QSO:  PKS\,0548$-$322 (see also \cite{Ja94}), 0959$-$443, 1355$-$41 (two
measurements with the same position angle also support the intrinsic nature of p here), 
1451$-$375 (these low
values agree with previously reported measurements -- see \cite{IT90}), 
PKS\,1514$-$24 (AP\,Librae, known to have shown high polarization -- see the 
references above), and PKS\,2115$-$30.

As expected from previous statistical studies, most of the objects with high, variable
polarization have flat radio spectra ($\alpha < 0.5$).  In Unified Schemes, this
means a near pole-on view of the AGN, in which the emission from the flat spectrum,
relativistically beamed compact core dominates that from the steep spectrum,
unboosted, extended lobes.  Similarly we expect all steep radio-spectrum AGN to 
show low
polarization.  There is one clear exception -- PKS 1452$-$217 with $\alpha \sim 0.70$\ 
and p $\sim 12$\%.   Meagre data (NED) suggest that the radio core is not variable, and
is completely resolved on VLBI scales (\cite{Pr85}).  
This may be a new example of the rare class of radio-loud quasar with
high polarization arising as a result of scattering, rather than synchrotron
emission from a blazar jet.  The classic case is OI\,287 (0752+258)
(Rudy \& Schmidt 1988; Goodrich \& Miller 1988; Antonucci, Kinney, \&
Hurt 1993), but we know of only three other examples (Hines \& Wills
1993; Hines et al. 1998; De Breuck et al. 1998; Brotherton et al. 1998).
The steep radio-spectrum suggests an edge-on view.  A
dusty disk may partially obscure unpolarized light reaching us directly
from the central continuum and broad emission line region, allowing
scattered, polarized light to dominate.  This hypothesis would be
consistent with its faintness (18.6$^{\rm m}$) at UV rest wavelengths.

It is also expected that a significant polarized component at short UV rest wavelengths
(higher redshifts) would require an exceptionally strong synchrotron component, i.e. one
that dominates the `normal' QSO continuum and broad emission lines.  This is because,
in radio-selected AGN,
the {\it optical} synchrotron spectrum is generally steep (F$_\nu \propto {\nu}^{-2}$)
compared with the normal (unpolarized) optical spectrum with $\alpha \sim $0.5 -- 1
(\cite{Wi91}; \cite{Wi92}).  Most AGN in Table\,1 
have low redshifts
(z $\lesssim 1$),
but PKS\,0046$-$315 is exceptional with z $= 2.721$\ and p $\sim$13\%.  It is also
one of the more optically luminous QSOs known (M$_{\rm abs} = -29.5$, \cite{VCV93}).
This led us to examine the redshift determination, but it appears to
be quite secure (\cite{Wi83}; \cite{Pe76}).  In strong support of
the presence
of a significant synchrotron continuum is the reported large amplitude, rapid variability
of 1$^m$\ over 2 months in 1975 (\cite{Pe76}).  Thus PKS\,0046$-$315 appears to
be the highest redshift blazar found so far.  However, we note that PKS\,0438$-$43 has 
z = 2.852, but the maximum polarization is significantly less, p$\sim$4.7\% $\pm 1$\%
(\cite{IT90}).

\acknowledgments

We thank Bruce Peterson for help with these observations, and David Jauncey for
encouragement to carry them out.  We also thank Mike Sitko for helpful refereeing.
BJW is grateful for support from NASA, under Long Term Space Astrophysics
grant NAG5-3431.
This research has made use of the NASA/IPAC Extragalactic Database (NED) which 
is operated by the Jet Propulsion Laboratory, California Institute of Technology, 
under contract with the National Aeronautics and Space Administration. 

\clearpage

\begin{deluxetable}{lllllrlrrr}
\footnotesize
\tablecaption{Polarization Measurements \label{tbl-1}}
\tablewidth{0pt}
\tablehead{
\colhead{Name} & \colhead{Original Name} & \colhead{Other Name} & \colhead{Class}& \colhead{z}&
\colhead{$\alpha$\tablenotemark{a}} & 
\colhead{UT Date}  & \colhead{p\tablenotemark{b} \,(\%)}   & \colhead{$\theta$($^\circ$)} & 
\colhead{isp\tablenotemark{c} \,(\%)}
}
\startdata

0046$-$315    &PKS            && QSO    & 2.721  &   0.00& 1978 12 27 & 13.3$\pm$ 2.0 & 159 & 0.0 \nl
0301$-$243    &PKS            && BLL HP &\nodata &   0.47& 1978 12 27 & 11.7\phs 1.0 & 57 & 0.0   \nl
0422$-$380    &PKS            && QSO    & 0.782  &$-$0.82& 1978 12 27 &  6.2\phs 3.0 &173 & 0.3  \nl
0448$-$392    &PKS            && QSO    & 1.288  &   0.00& 1978 12 27 &  2.9\phs 1.0 & 49 & 0.0   \nl
0521$-$365    &PKS\,0521$-$36 && BLL HP &        &   0.39& 1978 12 27 &  6.4\phs 0.5 &145 & 0.1   \nl
0537$-$441    &PKS            && BLL HP & 0.896  &$-$0.05& 1978 12 27 &  3.4\phs 0.5 &160 & 0.2   \nl
0548$-$322    &PKS            && BLL HP & 0.069  &   0.54& 1978 12 27 &  2.4\phs 0.8 & 24 & 0.0   \nl
              &               &&        &        &       & 1979 05 20 &  2.5\phs 0.5 & 78 &      \nl 
0629$-$418    &PKS            && QSO    & 1.416  &$-$0.54& 1979 05 19 &  3.4\phs 1.0 &177 & 0.5  \nl 
0637$-$752    &PKS\,0637$-$75 && QSO    & 0.654  &$-$0.51& 1978 12 27 &  1.9\phs 0.3 & 15 & 1.0  \nl 
0735+178      &PKS\,0735+17   && BLL HP &$>$0.424&$-$0.14& 1979 05 20 & 15.3\phs 0.7 & 33 & 0.3  \nl 
0754+100      &PKS  &OI\,+090.4& BLL HP & 0.66   &$-$0.24& 1978 12 27 &  9.5\phs 0.5 & 55 & 0.0  \nl 
              &               &&        &        &       & 1979 05 20 &  4.8\phs 0.3 & 66 &      \nl 
0808+019      &PKS            && BLL HP &\nodata &   0.09& 1979 05 19 &  8.8\phs 0.5 &127 & 0.2  \nl 
0818$-$128    &PKS            && BLL HP &\nodata &   0.02& 1979 05 19 & 18.2 0.5 & 90 & 0.7  \nl 
0823+033      &PKS            && BLL HP & 0.506  &$-$0.73& 1979 05 19 &  0.3 0.3 & 86 & 0.3  \nl 
0823$-$223    &PKS            && BLL HP &$>$0.910&$-$0.27& 1978 12 27 &  8.5 0.5 &122 & 0.6: \nl 
              &               &&        &        &       & 1979 05 20 & 12.7 0.4 &128 &      \nl 
0829+046      &PKS            && BLL HP & 0.18   &$-$0.07& 1978 12 27 & 10.5 0.6 & 26 & 0.1  \nl 
              &               &&        &        &       & 1979 05 19 & 20.5 0.7 & 58 &      \nl 
0851+202      &OJ\,287        && BLL HP & 0.306  &   0.42& 1978 12 27 & 12.2 1.0 & 76 & 0.2  \nl 
0858$-$771    &PKS\,0858$-$77 && QSO    & 0.490  &   0.39& 1978 12 27 &  2.4 0.6 &119 & 1.2  \nl 
0959$-$443    &PKS            && QSO    & 0.837  &   0.08& 1978 12 27 &  2.3 0.7 &174 & 0.4: \nl 
1020$-$103    &PKS            && QSO    & 0.197  &   0.43& 1979 05 19 &  0.4 0.4 &111 & 0.4  \nl 
1034$-$293    &PKS            && QSO HP & 0.312  &$-$0.20& 1979 05 19 & 15.3 0.3 &148 & 0.3  \nl 
1034$-$374    &PKS            && QSO    & 1.821  &   0.33& 1978 05 19 &  4.2 0.5 & 62 & 0.5  \nl %
1101$-$325    &PKS            && QSO    & 0.354  &   0.39& 1978 12 27 &  2.3 0.5 & 35 & 0.9  \nl 
1155+169      &PKS            && QSO    & 1.05   &   0.08& 1979 05 19 &  0.3 0.4 & 65 & 0.2  \nl 
1204$-$126    &PKS\,1204$-$12 && QSO    &\nodata &   0.77& 1979 05 19 &  3.0 1.0 & 69 & 0.4  \nl 
1206$-$238    &PKS\,B1206$-$238 && gal? &\nodata &$-$0.60& 1979 05 20 &  8.6 1.0 & 24 & 0.7  \nl 
1207$-$399    &PKS            && QSO    & 0.966  &   0.17& 1979 05 19 &  4.2 0.8 & 82 & 1.0  \nl 
1216$-$010    &PKS            && QSO    & 0.415  &$-$0.63& 1979 05 19 &  6.9 0.8 &  8 & 0.2  \nl 
1243$-$412    &PKS            && QSO    &\nodata &   0.59& 1979 05 20 &  2.6 1.0 & 16 & 1.1  \nl 
1256$-$220    &PKS\,1256$-$22 && QSO    & 1.306  &$-$0.59& 1979 05 20 &  5.2 0.8 &160 & 0.6  \nl 
1301$-$192    &PKS            && QSO    &\nodata &   0.81& 1979 05 20 &  2.0 0.5 & 79 & 0.5  \nl 
1302$-$102    &PKS            && QSO    & 0.286  &$-$0.06& 1979 05 19 &  1.0 0.4 & 70 & 0.2  \nl 
1307+121      &PKS\,1307+12C & 4C\,12.46& BLL    &\nodata &$-$1.08& 1979 05 20 &  5.4 0.5 & 20 & 0.0  \nl 
1349$-$439/R  &PKS            && BLL HP &\nodata &$-$0.55& 1979 05 19 & 21.5 0.5 &179 & 0.8  \nl 
1355$-$416    &PKS 1355$-$41  && QSO    & 0.313  &   0.89& 1979 05 19 &  1.8 0.4 & 90 & 0.6  \nl 
              &               &&        &        &       & 1979 05 20 &  2.4 0.6 & 88 &      \nl 
1400+162      &PKS\,1400+16   && BLL HP & 0.245  &   0.10& 1979 05 19 & 14.2 0.6 & 92 & 0.0  \nl 
              &               &&        &        &       & 1979 05 20 & 16.2 0.8 & 99 &      \nl 
1451$-$375    &PKS            && QSO    & 0.314  &$-$0.32& 1979 05 19 &  2.2 0.5 & 73 & 0.6  \nl 
              &               &&        &        &       & 1979 05 20 &  2.6 0.6 & 80 &      \nl 
1452$-$217    &PKS            && QSO    & 0.773  &   0.70& 1979 05 19 & 12.4 1.5 & 60 & 0.8  \nl 
1514+197      &PKS            && BLL HP &\nodata &$-$0.21& 1979 05 20 & 16.8 3.0 & 65 & 0.4  \nl 
1514$-$241    &PKS\,1514-24 &AP Librae & BLL HP & 0.048  &   0.14& 1979 05 19 &  1.1 0.2 &  9 & 1.5  \nl 
              &               &&        &        &       & 1979 06 24 &  3.1 0.1 &  3 &      \nl 
1532+016      &PKS 1532+01    && QSO HP & 1.435  &   0.26& 1979 05 19 &  7.9 1.0 &175 & 0.5  \nl 
1912$-$549    &PKS            && QSO    & 0.398  &   0.70& 1979 05 19 &  2.8 0.8 & 54 & 0.4  \nl 
              &               &&        &        &       & 1979 05 20 &  2.9 0.6 & 67 &      \nl 
1921$-$293\tablenotemark{d}    &PKS\,B1921-293 &OV$-$236       & BLL HP & 0.352  &$-$1.01& 1979 05 19 & 20.9 0.4 & 98 & 1.0  \nl 
              &               &&        &        &       & 1979 06 14 & 13.7 0.2 & 61 &      \nl 
1954$-$388    &PKS            && QSO HP & 0.630  &   0.00& 1979 05 20 &  6.5 1.5 &162 & 0.8  \nl 
2115$-$305    &PKS\,2115$-$30 && QSO    & 0.980  &   0.84& 1979 05 19 &  3.4 0.4 & 67 & 0.6  \nl 
2128$-$123    &PKS\,2128$-$12 && QSO    & 0.501  &$-$0.04& 1979 05 19 &  1.9 0.4 & 64 & 0.4  \nl 
2135$-$147    &PKS\,2135$-$14 && QSO    & 0.200  &   0.75& 1979 05 19 &  1.1 0.4 &100 & 0.4  \nl 
2141+175      &OX\,169        && QSO    & 0.213  &$-$0.49& 1979 05 19 &  1.3 0.4 & 66 & 1.0  \nl 
2145+067      &PKS\,2145+06   && QSO    & 0.990  &$-$0.39& 1979 05 19 &  1.0 0.6 &129 & 0.3  \nl 
2155$-$304    &PKS            && BLL HP & 0.116  &   0.15& 1979 05 19 &  4.5 0.1 & 31 & 0.0  \nl 
              &               &&        &        &       & 1979 06 14 &  6.1 0.2 & 32 &      \nl 
2216$-$038    &PKS\,2216$-$03 && QSO    & 0.901  &$-$0.73& 1979 05 19 &  0.8 0.7 &116 & 0.5  \nl 
2223$-$052    &PKS\,2223$-$05, 3C\,446        && QSO HP & 1.404  &   0.08& 1979 05 19 &  9.6 0.6 &161 & 0.5  \nl 
\enddata


\tablenotetext{a}{The radio spectral index, $\alpha$, between 2.7\,GHz and 5\,GHz is defined by
F$_{\nu} \propto {\nu}^{-\alpha}$.}
\tablenotetext{b}{No correction has been applied for positive bias in p (see \S2).}
\tablenotetext{c}{This is an upper limit to the degree of the interstellar polarization in
our Galaxy (see \S2).}
\tablenotetext{d}{We classify 1921-293 (OV$-$236) as a BL Lac object, rather than QSO as given in the literature,
because the optical spectrum showed only narrow emission lines (\cite{Wi81}).}

\end{deluxetable}

\clearpage



\begin{thebibliography}{}
\bibitem[Antonucci et al. 1993]{AnKiHu93} Antonucci, R. R. J., Kinney, A. L. \& Hurt, T.
   1993 \apj, 414, 506
\bibitem[Antonucci \& Ulvestad 1985]{AU85} Antonucci, R. R. J. \& Ulvestad, J. S. 1985, 
  \apj, 294, 158
\bibitem[Brand 1985]{Br85} Brand, P. W. J. L. 1985, in Proc. ``Active galactic nuclei'', 
  Manchester, England and Dover, NH, Manchester University Press, 1985,  215
\bibitem[Brindle et al. 1986]{Br86} Brindle C., Hough J. H., Bailey J. A., Axon D. J., 
  Hyland A. R. 1986, MNRAS, 221, 739
\bibitem[Brotherton et al. 1998]{BroWiDe98} Brotherton, M. S., Wills, B. J., Dey, A., 
  van Breugel, W., \& Antonucci, R. R. J.  1998, \apj, in press
\bibitem[Burstein \& Heiles 1982]{BH82} Burstein, D., \& Heiles, C. 1982, \aj, 87, 1165
\bibitem[De Breuck et al. 1998]{deBBroTra98} De Breuck, C., Brotherton, M. S., Tran, H. D.,
  van Breugel, W., \& Rottgering, H. J. A. 1998, \aj, in press
\bibitem[Gabuzda, Sitko \& Smith 1994]{GaSiSm96} Gabuzda, D. C., Sitko, M. L.,\& Smith, P. S. 
  1996, AJ, 112, 1877
\bibitem[Goodrich \& Miller 1988]{GoMi88} Goodrich, R. W., \& Miller, J. S. 1988, ApJ, 331, 332
\bibitem[Hines \& Wills 1993]{HiWi93} Hines, D. C. \& Wills, B. J. 1993, ApJ, 415, 82
\bibitem[Hines et al. 1998]{HiScWi98} Hines, D., Schmidt, G. D., Wills, B. J., Smith, P. S.
 \& Sowinski, L. G. 1998, \apj, submitted
\bibitem[Impey \& Tapia 1990]{IT90} Impey, C. D. \& Tapia, S. 1990, \apj, 354, 124 
\bibitem[Jannuzi, Smith, \& Elston 1994]{Ja94} Jannuzi, B.T., Smith, P. S, Elston, R. 1994,
   \apj, 428, 130
\bibitem[Kidger et al. 1995]{Ki95} Kidger, M. R., Gonzalez-Perez, J. N., De Diego, J. A.,
   Zapatero-Osorio, M. R., Hammersley, P. L., Cepa, J., DeVaney, N., Sahu, K., Vidal, I. 
   1995, A\&AS, 113, 431
\bibitem[Mead et al. 1990]{Me90} Mead, A. R. G., Ballard, K. R., Brand, P. W. J. L., 
  Hough, J. H., Brindle, C., \& Bailey, J. A.  1990, A\&AS, 83, 183
\bibitem[Moore et al. 1982]{Mo82} Moore, R. L., \etal\ 1982, \apj, 260, 415
\bibitem[Moore \& Stockman 1984]{MS84} Moore, R. L., \& Stockman, H. S 1984, \apj, 279, 465
\bibitem[Peterson et al. 1976]{Pe76} Peterson, B. A., Jauncey, D. L., 
  Wright, A. E., \& Condon, J. J.  1976, \apjl, 207, L5
\bibitem[Preston et al 1985]{Pr85} Preston, R. A., Morabito, D. D., Williams, J. G., 
  Faulkner, J., Jauncey, D. L., \& Nicolson, G.  1985, AJ, 90, 1599
\bibitem[Rudy \& Schmidt 1988]{RuSc88} Rudy, R. \& Schmidt, G. D. 1988, 331, 325
\bibitem[Rusk 1990]{Ru90} Rusk, R. 1990, JRASC, 84, 199
\bibitem[Scarpa \& Falomo 1997]{SF97} Scarpa, R. \& Falomo, R. 1997, A\&A, 325, 109
\bibitem[Serkowski, Mathewson, \& Ford 1975]{SMF75} Serkowski K., Mathewson, D. S., \& 
   Ford, V. L. 1975, \apj, 196, 261 
\bibitem[Sikora et al. 1997]{Si97} Sikora, M., Madejski, G., Moderski, R., 
   Poutanen, J. 1997, \apj, 484, 108
\bibitem[Stickel, Meisenheimer, \& K\"uhr 1994]{St94} Stickel, M., Meisenheimer, K., \& 
   K\"uhr, H. 1994, A\&A Suppl., 105, 211
\bibitem[Stockman, Moore, \& Angel 1984]{SMA84} Stockman, H. S., Moore, R. L., \& 
   Angel, J. R. P. 1984, \apj, 279, 485
\bibitem[Tran et al. 1998]{Tr98} Tran, H., Yuan, M. J., Wills, B. J., \& Wills, D.  1998,
    in preparation
\bibitem[V\'eron-Cetty \& V\'eron 1993]{VCV93} V\'eron-Cetty, M.-P. \& V\'eron, P. 1993, 
A Catalogue of Quasars \& Active Galactic Nuclei (ESO Sci. Rep., 13)(Garching: ESO)
\bibitem[Visvanathan 1972]{Vi72} Visvanathan, N. 1972, \pasp, 84, 248
\bibitem[Wagner 1996]{Wa96} Wagner, S. J. 1996, A\&A Suppl. 120, 495
\bibitem[Wardle \& Kronberg 1974]{WK74} Wardle, J. F. V., \& Kronberg, P. P. 1974, \apj, 194, 249
\bibitem[Wilkes et al. 1983]{Wi83} Wilkes, B.J., Wright, A. E., Jauncey, D. L., 
   Peterson, B. A.  1983, Publ. ASA, 5, 2
\bibitem[Wills 1991]{Wi91} Wills, B.J.  1991, in Proc. Symposium ``Variability of AGN'',
   held in May 1990 in Atlanta, Georgia, eds. Miller, R. \& Wiita, P.,  87
\bibitem[Wills et al. 1992]{Wi92} Wills, B.J., Wills, D., Antonucci, R.R.J., Barvainis, R.,
   \& Breger, M.  1992, \apj, 398, 454
\bibitem[Wills \& Wills 1981]{Wi81} Wills, D., \& Wills, B.J. 1981, Nature, 289, 384


\end{thebibliography}
\end{document}